\documentclass[a4paper]{article}
\newcommand{\be} {\begin{equation}}
\newcommand{\ee} {\end{equation}}
\newcommand{\bdm} {\begin{displaymath}}
\newcommand{\edm} {\end{displaymath}}
\newcommand{\bc} {\begin{center}}
\newcommand{\ec} {\end{center}}
\newcommand{\beqa} {\begin{eqnarray}}
\newcommand{\eeqa} {\end{eqnarray}}
\newcommand{\nn} {\nonumber}
\newcommand{\ra} {\rightarrow}
\newcommand{\lb} {\label}
\newcommand{\bfig} {\begin{figure}}
\newcommand{\efig} {\end{figure}}
\newcommand{\btab} {\begin{tabular}}
\newcommand{\etab} {\end{tabular}}
\newcommand{\hl} {\hline}
\usepackage{epsfig}
\begin{document}
\hfill\vbox{\hbox{OUTP 0160P}\hbox{MC-TH/01-13}}

\bigskip
 
\bc 
{\bf\Large RADIATIVE DECAYS OF EXCITED VECTOR MESONS}
\ec

\medskip

\bc
{\large F E Close}\\
{\large Department of Theoretical Physics, University of Oxford}\\
{\large Keble Road, Oxford, OX1 3NP, England}
\ec


\bc
{\large A Donnachie}\\
{\large Department of Physics and Astronomy, University of Manchester}\\
{\large Manchester M13 9PL, England}
\ec


\bc
{\large Yu S Kalashnikova}\\
{\large ITEP, Moscow, Russia}
\ec

\medskip

\begin{abstract}
\noindent Radiative decays of the $1^3S_1$ radial and $1^3D_1$ orbital excitations of 
the $\rho$, $\omega$ and $\phi$ are calculated in the quark model, using 
wave functions obtained variationally from the Hamiltonian with standard
quark-model parameters. The larger radiative widths should be measurable
at new high-intensity facilities being proposed, and in some cases may
be measurable in data from existing experiments. The radiative decays
are a strong discriminator between the $1^3S_1$ and $1^3D_1$ excitations,
and can also be used to provide unique information about the decay products.
\end{abstract}

\section{Introduction}

Radiative decays offer a rather direct probe of hadron structure. 
The coupling to the charges and spins of constituents reveals
detailed information about wave functions and can discriminate among models.
This can be particularly relevant in distinguishing gluonic
excitations of the $\pi,\rho$ (hybrids) from conventional excitations
with these overall $J^{PC}$. For example, in a hybrid $1^{--}$ the $q\bar{q}$
are in a spin-singlet, while for the $0^{-+}$ they are in a triplet: in each
case this is the reverse of what one is used to.

In this paper we investigate potential tests using radiative decays of
vector mesons to separate $q\bar q$ states from hybrids in the 1-2 GeV 
mass region, where light-flavour states are predicted to occur, and 
where {\it prima facie} candidates have been identified. Exotic $J^{PC} = 
1^{-+}$ signals have been reported around 1.4-1.6 GeV mass, the systematics 
of $1^{--}$ states in this region seem to point towards gluonic 
excitations being present and there are tantalising hints of unusual 
activity in the $0^{-+}$ partial wave in the 1.6-1.8 GeV region.

A quite separate problem is posed by the scalar mesons. It seems likely 
that the $a_0(980)$ and $f_0(980)$ are intimately linked to
the $K\bar K$ threshold with significant $q^2{\bar q}^2$ affinity. Then
the $q\bar q$ mesons are manifested in the 1.3 to 1.7 GeV region, presumably
mixed with the $0^{++}$ glueball of the lattice. A direct measure of their
flavour content would resolve this issue. We suggest that radiative
transitions from the $2S$ and $1D$ vector excitations can answer this.

The forthcoming generation of $e^+e^-$ facilities, such as the upgrade of 
VEPP at Novosibirsk, the use of initial state radiation (ISR) at BABAR to
study light-quark vectors and the proposed dedicated $e^+e^-$ collider
PEP-N, promise more than two orders of magnitude 
increase in data over present machines. There is a complementary 
high-intensity programme of photo- and electroproduction at Jefferson 
Laboratory. Together these will enable data on radiative decays to be 
obtained comparable in 
quality to that of present hadronic channels. Thus it is timely to ask 
whether, and under what circumstances, these data will be able to make
sensitive probes of radiative transitions and to isolate evidence for 
gluonic degrees of freedom. Radiative decays offer an alternative approach 
to resolving the various problems we have outlined. They are a much better 
probe of wave functions, and hence of models, than are hadronic decays. 
They can access final states which are kinematically excluded for hadronic 
decays. Historically, studies of light-quark radiative transitions 
have been restricted to the ground states. We extend these calculations to 
decays from the $2S$ and $1D$ excitations of the $\rho$, $\omega$ and $\phi$ 
to $1S$, $2S$ and $1P$ states, as these processes are now becoming accessible 
to experiment.

Our results are encouraging. We find that certain 
channels, with large partial widths, allow a clean separation of the $2S$ 
and $1D$ $q{\bar q}$ states, for example $f_2(1270)\gamma$ for the 
$\rho(1450)$, $f_1\gamma$ for the $\rho(1700)$, $f'_2(1525) \gamma$ for
the $\phi(1690)$ and $f_1(1420)$ for the $\phi(1900)$. Additionally the 
radiative decays of the $\rho(1700)$ and $\phi(1900)$ to $n\bar n$ 
and $s\bar s$ scalars respectively, probe the $f_0$ sector and provide 
a potential entr{\'e}e to flavour filtering.

In section 2 we review the present status of the interesting mesons in
the 1 to 2 GeV mass range.  We then turn in sections 3 and 4 to 
establishing our formalism for calculating radiative transition rates. 
Their magnitudes are computed in section 5, and in section 6 we identify 
a strategy for exploiting these results in forthcoming experiments.

\section{Mesons in the 1 to 2 GeV mass range}

The existence of two higher isovector vector mesons, the $\rho(1450)$ and
the $\rho(1700)$; their isoscalar counterparts, the $\omega(1420)$ and
$\omega(1650)$; and of an associated hidden-strangeness state, the
$\phi(1680)$, is well established \cite{PDG}.  Although there is general
consensus on the existence of these states, there is considerable disparity
on their masses and widths. Further, what is known about the composition of
their hadronic decays raises fundamental questions about the nature of these
states and our understanding of the mechanism of hadronic decays.

An apparently natural explanation for the higher-mass vector states is
that they are the first radial, $2^3S_1$, and first orbital, $1^3D_1$,
excitations of the $\rho$ and $\omega$ and the first radial excitation
of the $\phi$, as the generally-accepted masses \cite{PDG} are close
to those predicted by the quark model \cite{GI85}. However this argument is
suspect as the masses of the corresponding $J^P = 1^-$ strange mesons are
less than the predictions, particularly for the $2^3S_1$ at $1414 \pm 15$ MeV
\cite{PDG} compared to the predicted 1580 MeV \cite{GI85}. Quite apart from
comparing predicted and observed masses, one would expect the $n{\bar n}$
mesons to be 100 to 150 MeV lighter than their strange counterparts,
putting the $2^3S_1$ at less than 1300 MeV and the $1^3D_1$ below 1600 MeV.
Also this interpretation faces a further problem. The data on the $4\pi$
channels in $e^+e^-$ annihilation and $\tau$ decay are not compatible
with the $^3P_0$ model \cite{3P0,KI87,ABS96,BCPS97}, which works well for
decays of established ground-state mesons. For example widths predicted to
be large are found to be so; widths predicted to be small, are found to be
so; calculated widths agree with data to $25 - 40\%$; and signs of amplitudes
are predicted correctly. As far as one can ascertain the $^3P_0$ model is
reliable, but it has not been seriously tested for the decays of excited
states.

The $^3P_0$ model predicts that the decay of the isovector $2^3S_1$
to $4\pi$ is extremely small: $\Gamma_{2S \to a_1\pi} \sim 3$ MeV
and $\Gamma_{2S \to h_1\pi} \sim 1$ MeV, and other possible $4\pi$
decays are even smaller. For the isovector $1^3D_1$
the model predicts that the $a_1\pi$ and $h_1\pi$ decays are large and
equal: $\Gamma_{1D \to a_1\pi} \sim \Gamma_{1D \to h_1\pi} \sim 105$
MeV. All other possible $4\pi$ channels are small.
As $h_1\pi$ contributes only to the $\pi^+\pi^-\pi^0\pi^0$
channel in $e^+e^-$ annihilation, and $a_1\pi$ contributes to both
$\pi^+\pi^-\pi^+\pi^-$ and $\pi^+\pi^-\pi^0\pi^0$, then after subtraction
of the $\omega\pi$ cross section from the total $\pi^+\pi^-\pi^0\pi^0$,
one expects that
$\sigma(e^+e^- \to \pi^+\pi^-\pi^0\pi^0) > \sigma(e^+e^- \to
\pi^+\pi^-\pi^+\pi^-)$. This contradicts observation \cite{novonew} over
most of the available energy range, in which $\sigma(\pi^+\pi^-\pi^+\pi^-)
\approx 2\sigma(\pi^+\pi^-\pi^0\pi^0)$. Further, and more seriously, it
has been shown recently by the CMD collaboration at Novosibirsk
\cite{novonew} and by CLEO \cite{cleonew} that the dominant channel by far
in $4\pi$ (excluding $\omega\pi$) up to $\sim 1.6$ GeV is $a_1\pi$. This
is quite inexplicable in terms of the $^3P_0$ model. So the standard picture
is wrong for the isovectors, and there are serious inconsistencies in the
isoscalar channels as well. One possibility is that the $^3P_0$ model is
simply failing when applied to excited states, which is an intriguing
question in itself. An alternative is that there is new physics involved.

A favoured hypothesis is to include vector hybrids \cite{DK93,CP97,DK99}, 
that is
$q{\bar q}g$ states. The reason for this is that, firstly, hybrid states occur
naturally in QCD, and secondly, that in the relevant mass range the dominant
hadronic decay of the isovector vector  hybrid $\rho_H$ is believed to be
$a_1\pi$ \cite{CP97}. The masses of light-quark hybrids have been obtained in
lattice-QCD calculations \cite{LMBR97,Ber97,LS98,McN99,Mor00}, although
with quite large errors. Results from lattice QCD and other approaches,
such as the bag model \cite{BC82,CS83}, flux-tube models \cite{IP83},
constituent gluon models \cite{KY95} and QCD sum rules \cite{BDY86,LPN87},
show considerable variation from each other. So the absolute mass scale is
somewhat imprecise, predictions for the lightest hybrid lying between 1.3
and 1.9 GeV. However most models and lattice gauge calculations predict
the lightest hybrids to be at the upper end of this mass range. It does
seem generally agreed that the mass ordering is $0^{-+} < 1^{-+} < 1^{--}
< 2^{-+}$.

Evidence for the excitation of gluonic degrees of freedom has emerged in
several processes. A clear exotic $J^{PC}=1^{-+}$ resonance, the 
$\pi_1(1600)$, is seen \cite{E852c} in the $\eta'(958)\pi$ channel in the 
reaction $\pi^- N \to (\eta'(958))\pi N$.
Two experiments \cite{E852a,VESa} have evidence for this
exotic in the $\rho^0\pi^-$
channel in the reaction $\pi^- N \to (\pi^+\pi^-\pi^-) N$. A peak in the
$\eta\pi$ mass spectrum at $\sim 1400$ MeV with $J^{PC} = 1^{-+}$ in
$\pi^- N \to (\eta\pi^-) N$ has also been interpreted as a resonance
\cite{E852b}. Supporting evidence for the 1400 MeV state in the same mode comes
from ${\bar p}p \to \eta\pi^-\pi^+$ \cite{CB}. There is evidence \cite{VESb}
for two isovector $0^{-+}$ states in the mass region 1.4 to 1.9 GeV;
$\pi(1600)$ and $\pi(1800)$. The quark model predicts only one. Taking the
mass of the $1^{-+} \sim 1.4$ GeV and assuming the generally-agreed mass
ordering, then the $0^{-+}$ is at $\sim 1.3$ GeV
and the lightest $1^{--}$ at $\sim 1.65$ GeV, which is in the range required
for the mixing hypothesis to work. However if the $\pi_1(1600)$ is the 
lightest $1^{-+}$ state, so that the vector hybrids are comparatively
heavy, say $\sim 2.0$ GeV, then strong mixing with the radial and orbital 
excitations would be less likely. At this stage we should keep an open mind 
and consider both options.

The scalar mesons in the mass range 1.3 to 1.7 GeV provide another place
in which to look for gluonic hadrons as lattice calculations have now
become sufficiently stable to predict \cite{lat1}, in the quenched 
approximation, that the lightest glueball has $J^{PC}= 0^{++}$ and 
is in the mass range 1.45 to 1.75 GeV. 

There are more scalar mesons than the simple $q\bar q$ $1^3P_0$ nonet
can accomodate. The $f_0(980)$ and $a_0(980)$ mesons most probably do
not belong to this nonet \cite{AC95,Clo01}, and are either $q^2{\bar q}^2$ 
states or $K\bar K$ molecules. In any case they are associated with 
the nearby $K\bar K$ threshold. Then the possible candidates for the
$q\bar q$ $1^3P_0$ nonet are $a_0(1450)$, $K^*_0(1430)$, $f_0(1370)$,
$f_0(1500)$ and $f_0(1710)$. There is an obvious excess 
in the isoscalar-scalar sector, with the natural inference of there
being a glueball state present. 
 
The lightest scalar glueball should mix with the $q\bar q$ scalars in
the same mass region and recent studies \cite{lat2} on a coarse-grained
lattice suggest that such mixing is significant. While analyses 
\cite{AC95,CFL97,LW00,CK01}
of the mixing differ in some details, the conclusions exhibit common robust
features. The flavour content is predicted to have $n\bar n$ and $s\bar s$
in phase for the $f_0(1370)$ and $f_0(1710)$ (SU(3)-singlet tendency),
out of phase for the $f_0(1500)$ (SU(3)-octet tendency) and to have a 
glueball component in all three states. The detailed pattern of mixing 
was determined in \cite{CK01} by studying the complete set of decay
branching ratios into pseudoscalars \cite{Bar00} for the $f_0(1370)$, 
$f_0(1500)$ and $f_0(1710)$, and confirmed by comparing relative
production rates. The preferred scenario \cite{CK01} gives the bare
masses as $m_g =1443 \pm 24$ MeV, $m_{n{\bar n}} = 1377 \pm 10$ MeV
and $m_{s{\bar s}} = 1674 \pm 10$ MeV. Other solutions have been
found which have either a heavy glueball, $m_g > m_{s{\bar s}}$, or a
light glueball, $m_g < m_{n{\bar n}}$, and although less consistent
with the data they cannot be ruled out completely. The preferred 
solution is consistent with what one would expect naively from the 
$s{\bar s}-n{\bar n}$ mass difference of about 300 MeV, and places 
the glueball at the lower end of the mass range given by the lattice 
calculations.

The mixing scheme implies that the isovector partner of the $n{\bar n}$
state, the $a_0$, should have a mass of about 1400 MeV. There is an
indication that this state has been observed \cite{a0}. Any confirmation
of this controversial $a_0(1450)$ is of paramount importance, not only
for the problem of glueball-quarkonia mixing, but for the nature of the 
$0^{++}$ mesons.

The emerging data suggest that gluonic excitations, both for hybrids and 
glueballs, are rather lighter than quenched-lattice predictions and that
their effects will be apparent in the 1 to 2 GeV mass range. As we shall
see, radiative transitions can shed new light on the matter. 

\section{Radiative Decays of Quarkonia}

The initial meson $A$, with mass $m_A$, decays at rest to the final-state
meson $B$, with mass $m_B$ and a photon with three-momentum ${\bf p}$.
In the non-relativistic quark model the standard expression for the
transition amplitude has the form
\be
{\bf M}_{A\to B} = {\bf M}^{q}_{A\to B}+{\bf M}^{\bar q}_{A\to B}
\label{Mdef}
\ee
where ${\bf M}^{q}_{A\to B}$ and ${\bf M}^{\bar q}_{A\to B}$  describe the
emission of the photon from the quark and antiquark respectively:
\beqa
{\bf M}^q_{A\to B} &=& {{I_q}\over{2m_q}}\int d^3k
\big[Tr\{\phi^{\dagger}_
B(k-{\textstyle{{1}\over{2}}}p)\phi_A(k)\big\}(2{\bf k}-{\bf p})\nn\\
&&~~~~-iTr\{\phi_B^{\dagger}({\bf k}-{\textstyle{{1}\over{2}}}{\bf p})
{\bf \sigma}\phi_A({\bf k})\}\times{\bf p}\big]
\label{Mq}
\eeqa
and
\beqa
{\bf M}^{\bar q}_{A\to B} &=& {{I_{\bar q}}\over{2m_q}}\int d^3k
\big[Tr\{\phi_A({\bf k})\phi_B^{\dagger}({\bf k}+{\textstyle{{1}\over{2}}}
{\bf p})\}(2{\bf k}+{\bf p})\nn\\
&&~~~~-iTr\{\phi_A({\bf k})
{\bf\sigma}\phi_B^{\dagger}({\bf
k}+{\textstyle{{1}\over{2}}}{\bf p})\}\times{\bf p}\big]
\label{Mqbar}
\eeqa
where $I_q$ and $I_{\bar q}$ are isospin factors and $m_q$ is the quark mass.

We use matrix forms for the wave functions. For a meson $M$, with 
quark spin 0, total angular momentum $j$ and magnetic quantum number $m$
the wave function is given by
\be
\phi_M({\bf q}) = {{1}\over{\sqrt{2}}} {\hat{\bf 1}} Y_{jm}({\hat q})R_M(q)~.
\label{spin0}
\ee
For a meson with total angular momentum $j$, quark spin 1 and quark orbital 
momentum $l$, the corresponding wave function is
\be
\phi_M({\bf q}) = {{1}\over{\sqrt{2}}} {\bf Y}_{jlm}({\hat q}){\bf \sigma}
R_M(q)~.
\label{spin1}
\ee
Here ${\hat{\bf 1}}$ is the $2\times 2$ unit matrix and $\bf \sigma$ is the Pauli
matrix. The $R_M(q)$ are the mesonic radial wave functions in the momentum
representation

We calculate in the centre-of-mass of $e^+e^-$ annihilation as in that
case the virtual photon is polarised, and one can study angular 
distributions with respect to the beam direction. These are given in 
Appendix A. In Appendix B we show how to derive the helicity amplitudes
and their relation to the formalism used in the paper, which enables the
results to be transferred to any frame of reference.

The differential decay 
rate is evaluated for initial photon polarisation $m_0=1$ and is given by
\be
{{d\Gamma}\over{d\cos\theta}}= 4p{{E_B}\over{m_A}}\alpha I\sum|M_{A\to B}|^2
\label{rate}
\ee
where the sum is over final-state polarisations. In (\ref{rate}) $E_B$ is the
centre-of-mass energy of the final meson and $I = I^2_q = I^2_{\bar q}$ is the 
isospin factor. We consider the radiative decays of neutral vector mesons, so
the isospin factors for decays between $n{\bar n}$ or $s{\bar s}$ states are  
\beqa
I &=& {\textstyle{{1}\over{36}}}~~{\rm for~}n{\bar n} \to n{\bar n}~\rm{with~same~isospin}\nn\\
I &=& {\textstyle{{1}\over{4}}}~~{\rm for~}n{\bar n} \to n{\bar n}~\rm{with~different~isospin}\nn\\
I &=& {\textstyle{{1}\over{9}}}~~{\rm for~}s{\bar s} \to s{\bar s}.
\lb{iso1}
\eeqa
We take $\eta=\eta_8=(u{\bar u}+d{\bar d}-2s{\bar s})/\sqrt{6}$ and 
$\eta'=\eta_1=(u{\bar u}+d{\bar d}+s{\bar s})/\sqrt{3}$ so the isospin factors for
decays to $\eta$ and $\eta'$ are
\beqa
I &=& {\textstyle{{1}\over{108}}}~~{\rm for~}n{\bar n}~\rm{with~isospin~0} \to \eta\nn\\ 
I &=& {\textstyle{{1}\over{12}}}~~{\rm for~}n{\bar n}~\rm{with~isospin~1} \to \eta\nn\\
I &=& {\textstyle{{1}\over{54}}}~~{\rm for~}n{\bar n}~\rm{with~isospin~0} \to \eta'\nn\\
I &=& {\textstyle{{1}\over{6}}}~~{\rm for~}n{\bar n}~\rm{with~isospin~1} \to \eta'\nn\\
I &=& {\textstyle{{2}\over{27}}}~~{\rm for~}s{\bar s} \to \eta\nn\\
I &=& {\textstyle{{1}\over{27}}}~~{\rm for~}s{\bar s} \to \eta'.
\lb{iso2}
\eeqa
Radial wave functions are found variationally from the Hamiltonian
\be
H = \frac{p^2}{m_q} + \sigma r -\frac{4}{3}\frac{\alpha_s}{r} + C
\label{Ham}
\ee
with standard quark model parameters $m_q=0.33$ GeV for $u$ and $d$ quarks, 
$m_q=0.45$ GeV for $s$ quarks, $\sigma = 0.18$ GeV$^2$ and $\alpha_s=0.5$. 
The wave functions are taken to be Gaussian, that is of the form
$\exp(-k^2/(2\beta^2_M))$ multiplied by the appropriate polynomials, 
and $\beta$ treated as the variational parameter in (\ref{Ham}) for each 
of the $1S$, $1P$, $2S$ and $1D$ states. The resulting values of $\beta$ 
and the corresponding masses for $n\bar n$ states are given in table 1, and
those for $s\bar s$ states in table 2. 

\begin{table}
\bc
\btab{|c|c|c|c|c|}
\hl
& $M (GeV)$ & $\beta (GeV)$ \\
\hl
$1S$ & $0.700$ & $0.313$ \\
\hl
$1P$ & $1.262$ & $0.274$ \\
\hl
$2S$ & $1.563$ & $0.253$ \\
\hl
$1D$ & $1.703$ & $0.255$ \\
\hl
\etab
\ec
\bc
Table 1. Effective masses and corresponding $\beta$ from the variational
solution of (\ref{Ham}) for $n\bar n$ states.
\ec
\end{table}

\bigskip

\begin{table}
\bc
\btab{|c|c|c|c|c|}
\hl
& $M (GeV)$ & $\beta (GeV)$ \\
\hl
$1S$ & 1.000 & 0.355 \\
\hl
$1P$ & 1.527 & 0.307 \\
\hl
$2S$ & 1.793 & 0.285 \\
\hl
$1D$ & 1.932 & 0.285 \\
\hl
\etab
\ec
\bc
Table 2. Effective masses and corresponding $\beta$ from the variational
solution of (\ref{Ham}) for $s\bar s$ states.
\ec
\end{table}

\bigskip

It is well-known that the use of exact wave functions is necessary to
reproduce the low-energy theorems. 
For example, the expression for the $E1$ transition
amplitude (equations (\ref{Mq}) and (\ref{Mqbar}) in the $p=0$ limit) can be
written in the
dipole form using the relation $\vec {k} = im_q[H\vec {r}]/2$.
The amplitude is proportional to the overlap integral
\be
\frac{im_q}{2}(E_A-E_B)\int r^3dr\psi_A(r)\psi_B(r), 
\lb{dipole}
\ee
which immediately gives the threshold $p^3$ behaviour of the electromagnetic
width demanded by gauge invariance. This means that not only should the 
``correct'' values of the oscillator parameter $\beta$ be used, but also
the masses of the initial and final states should be taken to be ``correct''
eigenvalues of the quark model Hamiltonian (\ref{Ham}). One can
systematically improve the variational ansatz adopted for the wave
functions, using more sophisticated trial wave functions \cite{GI85,Bonnaz} 
or numerical
solutions \cite{Bonnaz}. Fortunately, in most cases in practice, the
``correct'' masses and the known 
physical masses do not differ by much, and the resulting differences in the
decay widths are smaller than other uncertainties in the calculation. However
this is not true of decays to the pseudoscalars $\pi$, $\eta$ and $\eta'$.

The constituent quark model in its naive form (\ref{Ham}) works reasonably
well, with the exception of the lowest pseudoscalars as the Goldstone-boson
nature of these particles is not naturally accomodated. More complicated 
versions,
such as \cite{GI85}, include relativistic kinematics and properly-smeared
spin-dependent terms, and the agreement with data on spectra is quite
remarkable. Similar results are obtained in the QCD string model
\cite{KNS}, where it is shown that the constituent masses of quarks appear
dynamically due to the QCD-inspired string-type interaction. The most
serious drawback of such a picture is that the issue of chiral symmetry
breaking is completely beyond its scope. It is possible to describe
the low-lying pion in the constituent quark models, but the Goldstone
nature of the pion is completely lost. However, the chiral properties can
be naturally incorporated into the constituent picture in the framework of
the Hamiltonian approach in the Coulomb gauge \cite{Swanson1}. In this
approach, a chirally-noninvariant vacuum is constructed, which implies the
existence of the Goldstone boson as the lowest $q \bar q$ state. The
axial current is conserved in the chiral limit, and all the relations of
current algebra are satisfied. Some results of this approach are relevant
to our purposes. First, even in the chiral limit, where the pionic wave
function has very peculiar properties, the other mesons behave to large
extent as quark model ones (for the details see \cite{Cotanch}). Second,
while the confining interaction alone can describe the observed $\pi$-$\rho$
splitting, the predictions for the quark condensate and pion decay
constant are too small, and can be improved by inclusion of the hyperfine
interaction \cite{Swanson2}. The net result of these studies is that
if the effective degrees of freedom are properly defined, then
the constituent quark model is a good approximation, with most important
ingredients absorbed in the quark model parameters.

This means that, as a first approximation, the mesonic wave functions can be
found from the Hamiltonian (\ref{Ham}). Also, for the pion, the
``correct'' $\beta$ should be larger than the $313$ MeV for $1S$ states
shown in table 1. We will return to this point below. 

\section{Radiative Decay Widths of Vector Mesons}

Expressions for the full angular distributions for the radiative decay
widths of the $1^3S_1$, $2^3S_1$ and $1^3D_1$ neutral vector mesons
are given in Appendix A. Here we give only the total widths. In these 
results we define
\be
\beta^2 = \frac{2\beta^2_A\beta^2_B}{(\beta^2_A + \beta^2_B)}
\lb{beta}
\ee
and
\be
\lambda = \frac{\beta^2_A}{2(\beta^2_A + \beta^2_B)}.
\lb{lambda}
\ee

A method for deriving the following expressions when $\beta_A = \beta_B$
is given in Appendix B. The generalisation to arbitrary $\beta_A$ and 
$\beta_B$ is straightforward but algebraically tedious.

${\bf 1^3S_1 \ra 1^1S_0}$
\be
\Gamma^0_S=\frac{4}{3}\alpha p\frac{E_B}{m_A}\frac{p^2}{m^2_q}IF^2_{S1},
\lb{S11}
\ee

${\bf 2^3S_1 \ra 1^1S_0}$
\be
\Gamma = \frac{3}{2}\left(\Big(\frac{\beta^2}{\beta_A^2}-1\Big)+
\frac{2\lambda^2p^2}{3\beta_A^2} \right)^2\Gamma^0_S.
\lb{S12}
\ee

${\bf 2^3S_1 \ra 2^1S_0}$
\be
\Gamma = \frac{9}{4}
\left(\Big(\frac{5\beta^4}{3\beta_A^2\beta_B^2}-1\Big)+\frac{4\lambda^2p^2}
{3\beta^2}
\Big(\frac{\beta^4}{3\beta_A^2\beta_B^2}-1\Big)+\frac{4\lambda^2p^4}
{9\beta_A^2\beta_B^2} \right)^2\Gamma^0_S
\lb{S13}
\ee
In (\ref{S11}), (\ref{S12}) and (\ref{S13}) $F_{S1}$ is defined by
\be
F_{S1}=\frac{\beta^3}{\beta_A^{3/2}\beta_B^{3/2}}\exp\Big(-\frac{p^2}
{8(\beta_A^2+\beta_B^2)}\Big).
\lb{FS1}
\ee

${\bf 2^3S_1 \rightarrow 1^3P_0}$
\be
\Gamma = \frac{16}{27}\alpha p\frac{E_B}{m_A}\frac{\beta^2}{m_q^2}
\Big(G_E^2 - \frac{p^2}{2\beta^2}G_EG_M+\frac{1}{16}\Big(\frac{p^2}{\beta^2}
\Big)^2G_M^2\Big)IF^2_{S2},
\lb{S21}
\ee

${\bf 2^3S_1 \rightarrow 1^3P_1}$
\be
\Gamma = \frac{16}{9}\alpha p\frac{E_B}{m_A}\frac{\beta^2}{m_q^2}
\Big(G_E^2 + \frac{p^2}{4\beta^2}G_EG_M+\frac{1}{32}\Big(\frac{p^2}{\beta^2}
\Big)^2G_M^2\Big)IF^2_{S2},
\lb{S22}
\ee

${\bf 2^3S_1 \rightarrow 1^3P_2}$
\be
\Gamma = \frac{80}{27}\alpha p\frac{E_B}{m_A}\frac{\beta^2}{m_q^2}
\Big(G_E^2 + \frac{p^2}{4\beta^2}G_EG_M+\frac{7}{160}\Big(\frac{p^2}{\beta^2}
\Big)^2G_M^2\Big)IF^2_{S2}.
\lb{S23}
\ee
In (\ref{S21}), (\ref{S22}) and (\ref{S23}) $F_{S2}$ is defined by
\be
F_{S2} = \frac{\beta^4}{\beta^{3/2}_A\beta^{5/2}_B} 
\exp{\Big(-\frac{p^2}{8(\beta^2_A + \beta^2_B)}\Big)},
\lb{FS2}
\ee
$G_E$ by
\be
G_E = \frac{5\beta^2}{2\beta^2_A} - \frac{3}{2}
+ \frac{\lambda^2p^2}{\beta^2_A},
\lb{GE}
\ee
and $G_M$ by
\be
G_M = \frac{4\lambda\beta^2}{\beta^2_A} + 
\Big(\frac{\beta^2}{\beta^2_A} - 1 \Big)(6\lambda - 3) + 
\frac{2\lambda^2p^2}{\beta^2_A}(2\lambda - 1).
\lb{GM}
\ee

${\bf 1^3D_1 \ra 1^1S_0}$
\be
\Gamma^0_D = \frac{8}{45}\alpha
p\frac{E_B}{m_A}\lambda^4\frac{p^4}{\beta^4}\frac{p^2}{m_q^2}IF^2_{D1},
\lb{D11}
\ee

${\bf 1^3D_1 \ra 2^1S_0}$
\be
\Gamma = \frac{3}{2}\left(\Big(\frac{2\lambda^2p^2}{3\beta_B^2}-1\Big)
+\frac{7\beta^2}
{3\beta_B^2} \right)^2\Gamma^0_D.
\lb{D12}
\ee
In (\ref{D11}) and (\ref{D12}) $F_{D1}$ is defined by
\be
F_{D1}=\frac{\beta^5}{\beta_A^{7/2}\beta_B^{3/2}}\exp\Big(-\frac{p^2}
{8(\beta_A^2+\beta_B^2)}\Big).
\lb{FD1}
\ee

${\bf 1^3D_1 \ra 1^3P_0}$
\be
\Gamma = \frac{80}{27}\alpha p\frac{E_B}{m_A}\frac{\beta^2}{m_q^2}
\Big(1+\frac{p^2}{5\beta^2}\lambda(1+2\lambda)+
\frac{p^4}{10\beta^4}\lambda^2(-1+2\lambda)\Big)^2IF^2_{D2},
\lb{D21}
\ee

${\bf 1^3D_1 \ra 1^3P_1}$
\beqa
\Gamma &=& \frac{20}{9}\alpha p\frac{E_B}{m_A}\frac{\beta^2}{m_q^2}
\Big(1+\frac{p^2}{5\beta^2}\lambda(7+4\lambda)
+\frac{p^4}{50\beta^4}\lambda^2(5+68\lambda+8\lambda^2)\nn\\
&+&\frac{4p^6}{25\beta^6}\lambda^3(-2+3\lambda+2\lambda^2)
+\frac{2p^8}{25\beta^8}\lambda^4(1-4\lambda+4\lambda^2)\Big)IF^2_{D2},
\lb{D22}
\eeqa

${\bf 1^3D_1 \ra 1^3P_2}$
\beqa
\Gamma &=&\frac{4}{27}\alpha p\frac{E_B}{m_A}\frac{\beta^2}{m_q^2}
\Big(1+\frac{p^2}{5\beta^2}\lambda(17+4\lambda)\nn\\
&+&\frac{p^4}{10\beta^4}\lambda^2(83-316\lambda+520\lambda^2)
+\frac{22p^6}{5\beta^6}\lambda^3(1-6\lambda+8\lambda^2)\nn\\
&+&\frac{8p^8}{5\beta^8}\lambda^4(1-4\lambda+4\lambda^2)\Big)
IF^2_{D2}.
\lb{D23}
\eeqa
In (\ref{D21}), (\ref{D22}) and (\ref{D23}) $F_{D2}$ is defined by
\be
F_{D2}=\frac{\beta^6}{\beta_A^{7/2}\beta_B^{5/2}}
\exp\Big(-\frac{p^2}{8(\beta_A^2+\beta_B^2)}\Big)
\lb{FD2}
\ee

\section{Results}

We first give the numerical results for all relevant radiative decays.
Discussion of these results is in two parts. The first treats decays to 
$^3P_J$ states, with the emphasis on unique signatures for specific 
vector $q\bar q$ states. The second deals with radiative decays to scalars
as a flavour filter with implications for the scalar glueball mass. We
then discuss qualitatively hybrid meson radiative transitions.

\subsection{Numerical results}

A simple test of the validity of tables 1 and 2 is provided by the well-known 
decays $\rho \to \eta\gamma$, $\omega \to \eta\gamma$, $\phi \to \eta\gamma$ 
and $\phi \to \eta'\gamma$. The calculated widths in keV, using $\beta=0.313$ 
from table 1 for the $n\bar n$ decays and $\beta=0.355$ from table 2 for the 
$s\bar s$ decays but the physical masses in each case, are compared with the
experimental widths \cite{PDG} in table 3. The agreement for the $\rho$ and 
$\omega$ decays is clearly satisfactory, that for the decay 
$\phi \to \eta\gamma$ less so.

\bfig[t]
\bc
\begin{minipage}{90mm}
\epsfxsize90mm
\epsffile{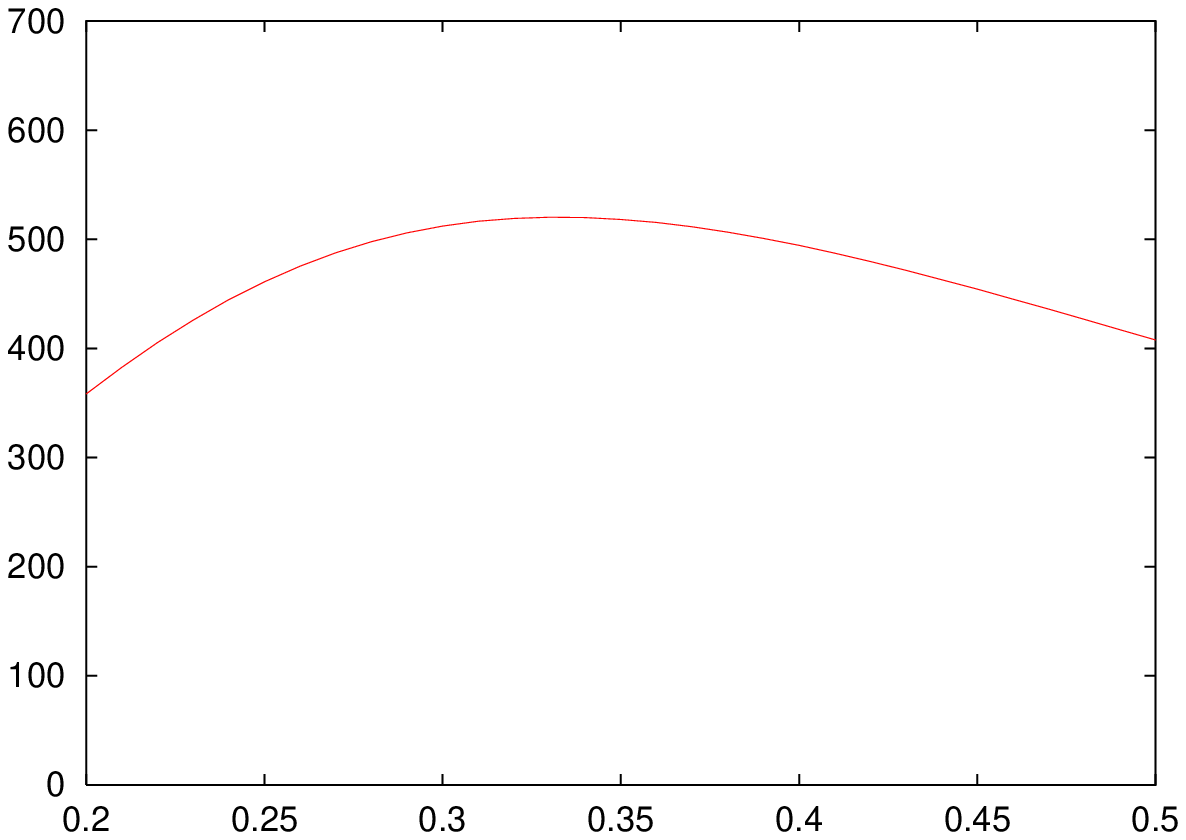}
\begin{picture}(0,0)
\setlength{\unitlength}{1mm}
\put(-9,62){$\Gamma_{\omega \to \pi\gamma}$}
\put(-8,57){(keV)}
\put(60,1){$\beta_\pi$~~(GeV)}
\end{picture}
\end{minipage}
\ec
\caption{Variation of the width for the decay $\omega \to \pi\gamma$ as a 
function of $\beta_\pi$ with $\beta_\omega$ fixed at 0.313 according to 
table 1.}
\lb{omegapi}
\efig

\bigskip

\begin{table}
\bc
\btab{|c|c|c|}
\hl
Decay& Model & Experiment\\
\hl
$\rho \to \eta\gamma$ & 39.6 & $36 \pm 13$\\
\hl
$\omega \to \eta\gamma$ & 4.7 & $5.5 \pm 0.9$\\
\hline
$\phi \to \eta\gamma$ & 92.7 & $57.82 \pm 1.53$\\
\hl
$\phi \to \eta'\gamma$ & 0.35 & $0.30 \pm 0.15$\\
\hl
\etab
\ec
\bc
Table 3. Radiative widths for $1^3S_1$ decays to $1^1S_0$.
\ec
\end{table}

We can use the decays $\rho \to \pi\gamma$ and $\omega \to \pi\gamma$, again 
with physical masses, to estimate the appropriate value of $\beta_\pi$ for 
decays to $\pi\gamma$. The variation of the width for $\omega \to \pi\gamma$ 
with $\beta_\pi$ is shown in figure 1. At no point does it reach the 
experimental value \cite{PDG} of $717 \pm 43$ keV. The maximum of the curve 
is 520 keV and occurs at $\beta_\pi = 0.335$, which we take to be the optimum 
value. The discrepancy between the model calculation and the experimental 
value is an indication of the uncertainty in the evaluation of decays to 
$\pi\gamma$.   

The radiative decay widths for the $n\bar{n}$ states are given in table 4. 
For purposes of illustration, we have assumed that the $f_0(1370)$ is a pure 
$n\bar n$ state and have ignored possible mixing in the $1^3P_0$ nonet. This 
is certainly an oversimplification as there is good evidence \cite{CK01},
discussed in section 2, for scalar mixing within this nonet and with a scalar
glueball. We return to this question below. The effect of phase space is
clearly seen in the deviations from the naive 9:1 ratios of (\ref{iso1}) and 
(\ref{iso2}). 
The radiative decay widths for the $s\bar s$ states are given in table 5.
As for the $n\bar n$ decays, we have ignored mixing and assumed that the 
$f_0(1710)$ is pure $s\bar s$. 

\begin{table}[t]
\bc
\btab{|c|c|c|c|c|}
\hl
& $\Gamma(\rho(1450))$ & $\Gamma(\omega(1420))$ & $\Gamma(\rho(1700))$ &
$\Gamma(\omega(1650))$ \\
\hl
$\pi\gamma$ & 61 & 510 & 3.8 & 40 \\
\hl
$\eta\gamma$ & 106  & 11 & 12 & 1.1 \\
\hl
$\eta'\gamma$ & 61 & 5.7 & 6 & 0.5 \\
\hl
$\pi(1300)\gamma$ & 5.9 & 29 & 0.4 & 1.7 \\
\hl
$\eta(1295)\gamma$ & 57 & 3.6 & 3.9 & 0.2 \\
\hl
$f_0(1370)\gamma$ & 64 & 4.7 & 899 & 88 \\
\hl
$a_0(1450)\gamma$ &  &  & 82 & 612 \\  
\hl
$f_1(1285)\gamma$ & 349 & 33 & 1097 & 106 \\
\hl
$a_1(1260)\gamma$ & 43 & 341 & 129 & 1016 \\
\hl
$f_2(1270)\gamma$ & 712 & 67 & 148 & 13 \\
\hl
$a_2(1320)\gamma$ & 59 & 413 & 13 & 91 \\
\hl
\etab
\ec
\bc
Table 4. Results for radiative decays of $n\bar n$ states in keV.
\ec
\end{table}


\begin{table}
\bc
\btab{|c|c|c|}
\hl
& $\Gamma(\phi(1680))$ & $\Gamma(\phi(1930))$ \\
\hl
$\eta\gamma$ & 94 & 9 \\
\hl
$\eta'\gamma$ & 21 & 1.8 \\
\hl
$\eta(1440)\gamma$ & 47 & 10 \\
\hl
$f_0(1710)\gamma$ & & 188 \\
\hl
$f_1(1420)\gamma$ & 148 & 408 \\
\hl
$f_2(1525)\gamma$ & 199 & 37 \\
\hl
\etab
\ec
\bc
Table 5. Results for radiative decays of $s\bar s$ states in keV.
\ec 
\end{table}

We have commented above on the uncertainties in the decays $\rho \to 
\pi\gamma$ and $\omega \to \pi\gamma$. The situation becomes even more 
uncertain for the decays $\rho(1450) \to \pi\gamma$ and $\omega(1420) 
\to \pi\gamma$. These decays proceed essentially via the spin-flip
part of the amplitude which vanishes in the nonrelativistic limit for
orthogonal wave functions. For Gaussian wave functions the amplitude is
proportional to $(\beta^2/\beta_A^2 - 1)+2\lambda^2p^2/(3\beta_A^2)$, see
(\ref{S12}), where the term $(\beta^2/\beta_A^2 - 1)$ measures the
nonorthogonality of the wave functions. Note that one should not expect 
the radial wave functions of the $2^3S_1$ and $1^1S_0$ states to be
orthogonal in the fully relativistic theory, as spin-dependent forces
are not treated as perturbations there. With $\beta_A \ne \beta_B$ the
resulting width is rather sensitive to the actual value of $\beta_B$.
The value of $\beta_\pi$ is not determined by the model, and as the
width for $\omega \to \pi\gamma$ varies rather slowly with $\beta_\pi$,
our choice of $\beta_\pi = 0.335$ has a rather large error. The effect
on the width of $\omega(1420) \to \pi\gamma$ of varying $\beta_\pi$ is
shown in \hbox{figure \ref{omegapi2}.}

\bfig
\bc
\begin{minipage}{90mm}
\epsfxsize90mm
\epsffile{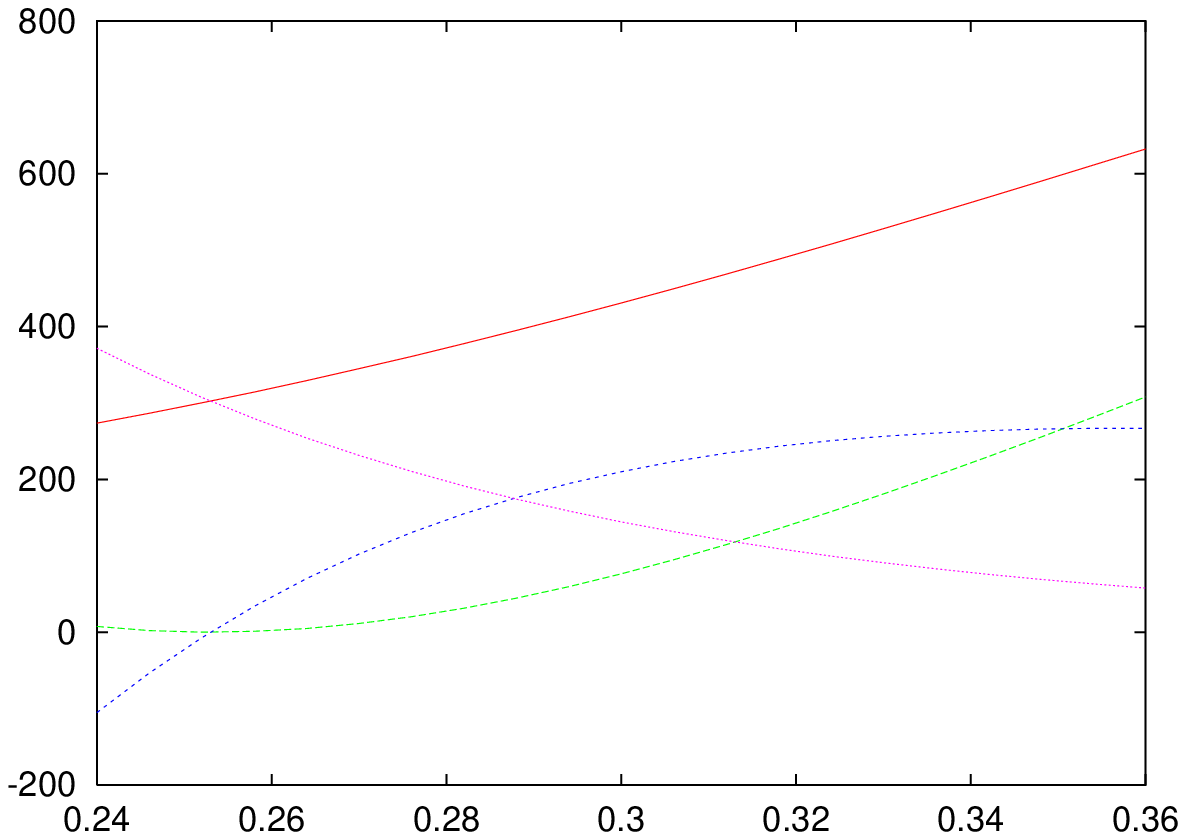}
\begin{picture}(0,0)
\setlength{\unitlength}{1mm}
\put(-12,60){$\Gamma_{\omega(1420) \to \pi\gamma}$}
\put(-8,55){(keV)}
\put(60,1){$\beta_\pi$~~(GeV)}
\end{picture}
\end{minipage}
\ec
\caption{Variation of the width for the decay $\omega(1420) \to \pi\gamma$ 
as a function of $\beta_\pi$. The solid line is the total width, the dashed 
line is the contribution from the first term in (\ref{S12}), the dotted line
is the contribution from the second term in (\ref{S12}) and the short-dashed
line is the contribution from the interference term.}
\lb{omegapi2}
\efig
\vskip 5truemm
\bfig
\bc
\begin{minipage}{90mm}
\epsfxsize90mm
\epsffile{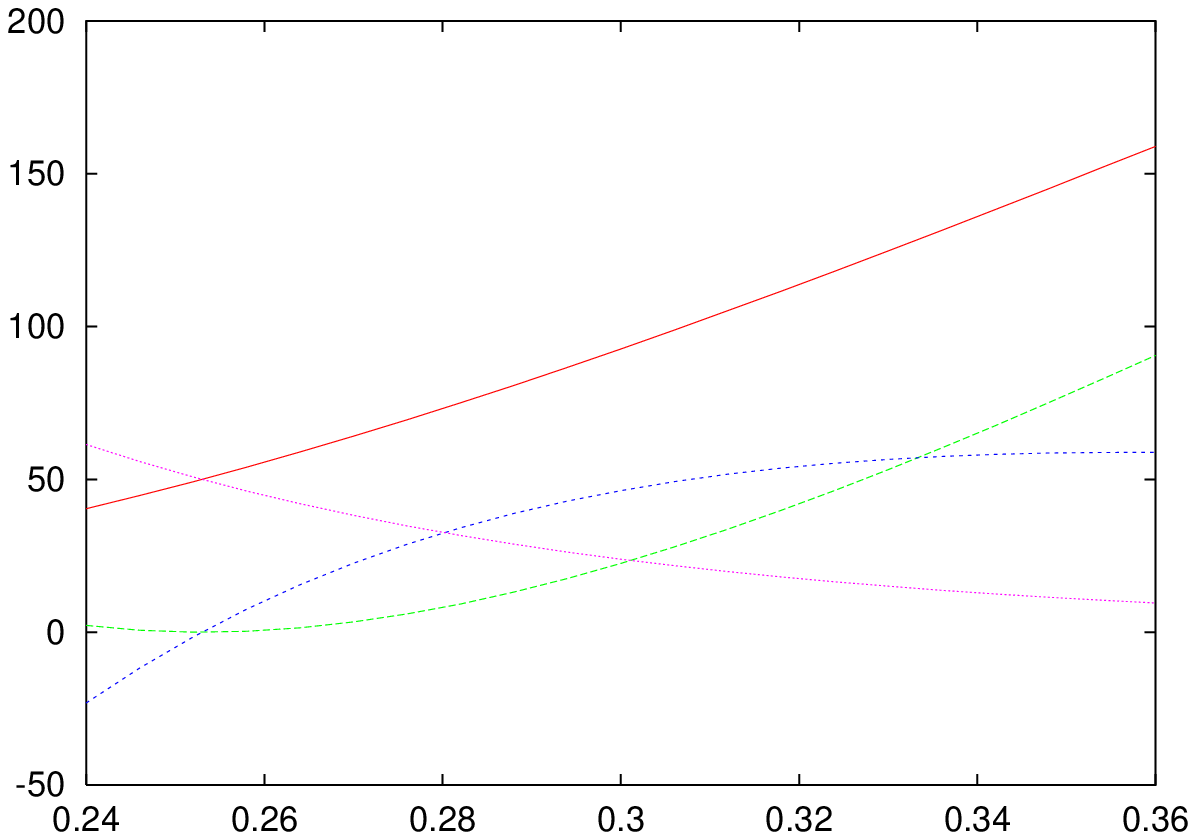}
\begin{picture}(0,0)
\setlength{\unitlength}{1mm}
\put(-12,60){$\Gamma_{\rho(1450) \to \eta\gamma}$}
\put(-8,55){(keV)}
\put(60,1){$\beta_\eta$~~(GeV)}
\end{picture}
\end{minipage}
\ec
\caption{Variation of the width for the decay $\rho(1450) \to \eta\gamma$ 
as a function of $\beta_\eta$. The solid line is the total width, the dashed 
line is the contribution from the first term in (\ref{S12}), the dotted line
is the contribution from the second term in (\ref{S12}) and the short-dashed
line is the contribution from the interference term.}
\lb{rhoeta}
\efig
 
The first term in (\ref{S12}) increases rapidly with increasing $\beta_\pi$,
and for $\beta_\pi$ in the range we are considering it is this term and the 
interference term which dominate the width. The direct contribution from the
second term in (\ref{S12}) becomes increasingly unimportant. A similar result
holds for the decays $\rho(1450) \to \eta\gamma$ and $\omega(1429) \to 
\eta\gamma$ as can be seen in figure \ref{rhoeta}. As a consequence we 
suggest that the errors on the $2S$ decays to $\pi\gamma$ and $\eta\gamma$
are of the order of $\pm 50\%$.

An estimate of the uncertainties for other decays can be obtained by looking 
at the contributions from terms in increasing powers, $n$, of $p^2/\beta^2$. 
These are given in table 6, as percentages, for some of the larger widths. 

\begin{table}
\bc
\btab{|c|c|c|c|c|c|}
\hl
&n=0&n=1&n=2&n=3&n=4\\
\hl
$\rho(1450) \to f_1(1285)\gamma$&106.2&$-6.4$&0.2&&\\
\hl
$\rho(1450) \to f_2(1270)\gamma$&93.1&6.6&0.3&&\\
\hl
$\rho(1700) \to f_0(1370)\gamma$&85.4&14.8&$-0.2$&$-0.1$&0.1\\
\hl
$\rho(1700) \to f_1(1285)\gamma$&56.5&39.6&4.7&$-0.9$&0.1\\
\hl
$\phi(1680) \to f_1(1420)\gamma$&112.7&$-13.5$&0.8&&\\
\hl
$\phi(1680) \to f_2(1525)\gamma$&95.0&4.8&0.2&&\\
\hl
$\phi(1900) \to f_0(1710)\gamma$&94.7&5.3&0.0&0.0&0.0\\
\hl
$\phi(1900) \to f_1(1420)\gamma$&52.1&43.2&6.0&$-1.4$&0.1\\
\hl
\etab
\ec
\bc
Table 6. Contributions, in percent, from terms in increasing powers, $n$,
of $p^2/\beta^2$.
\ec
\end{table}

\bigskip

The implications of table 6 are that the results for the decays of 
$\rho(1450)$ to $f_1(1285)\gamma$ and $f_2(1270)\gamma$, and the corresponding 
$\omega(1420)$ decays to $a_1(1260)\gamma$ and $a_2(1320)\gamma$ are very 
reliable, the decay of $\rho(1700)$ to $f_0(1370)\gamma$ is solid, but
the decay of $\rho(1700)$ to $f_1(1285)\gamma$, and the corresponding 
$\omega(1650)$ decay to $a_1(1260)\gamma$ should be treated with caution. 
This is not suprising because of the considerable increase in phase space. 
The situation is similar for the decays of $\phi(1690)$ and $\phi(1900)$,
with the $f_1(1420)\gamma$ decay of the latter being the only one with any
degree of uncertainty.

\subsection{Decays to $^3P_J$ states}

It is clear from table 4 that some decays provide very clear signatures
for particular excitations. Obvious ones are $\rho(1450) \to
f_2\gamma$ and $\omega(1420) \to a_2\gamma$; $\rho(1700) \to f_0\gamma$ 
and $\omega(1650) \to a_0\gamma$; $\rho(1700) \to f_1\gamma$ and 
$\omega(1650) \to a_1\gamma$. However, in $e^+e^-$ annihilation experiments, 
isovector states are produced at approximately nine times the rate of the
corresponding isoscalar states. Thus, for example, the effective $a_0\gamma$
rate from the decay of $\omega(1650)$ will be the same as that from 
$\rho(1700)$. Conversely the contamination of the $f_2\gamma$ decay of 
$\rho(1450)$ by the corresponding decay of $\omega(1420)$ will only be
at the $1\%$ level. The $f_2(1525)\gamma$ decay of the $\phi(1690)$ is
important as it provides a unique signature for the $s\bar s$ state, since
in contrast to hadronic decays it is unaffected by $\omega(1650)$ and 
$\phi(1690)$ mixing. As we argue in section 5.4 radiative decays of a 
hybrid $\rho_H$ to $f_2\gamma$ and of a hybrid $\phi_H$ to $f_2(1525)\gamma$
are strongly suppressed, so there is no ambiguity.

Although the $f_1(1420)\gamma$ decay of the $\phi(1900)$
looks like a clean signature of this state, it should be recalled that the
value quoted in table 5 should be treated with some caution. Despite this
uncertainty, the width is necessarily much larger than that for the 
$\eta(1440)\gamma$ decay. This provides a mechanism for producing the
$f_1(1420)$ without contamination from $\eta(1440)$ with which it shares
many common hadronic decay channels.

It should be noted that there is some uncertainty about the 
mass of the $\rho(1450)$ and $\omega(1420)$. For the isovector states, the
most extreme low mass comes from an analysis of of the $\pi^+\pi^-$
spectrum in the reaction $K^-p \to \pi^+\pi^-\Lambda$ \cite{LASS},
which gives $1266 \pm 14$ MeV. An equally low mass, $1250 \pm 29$ MeV
has been suggested \cite{SND} for the isoscalar channel from an analysis
of $e^+e^- \to \pi^+\pi^-\pi^0$. Given this uncertainty, we show the mass
variation of the width for the decay of the isovector radial ($2^3S_1$) 
excitation to $f_2\gamma$ in figure 2.

\bfig[t]
\bc
\begin{minipage}{90mm}
\epsfxsize90mm
\epsffile{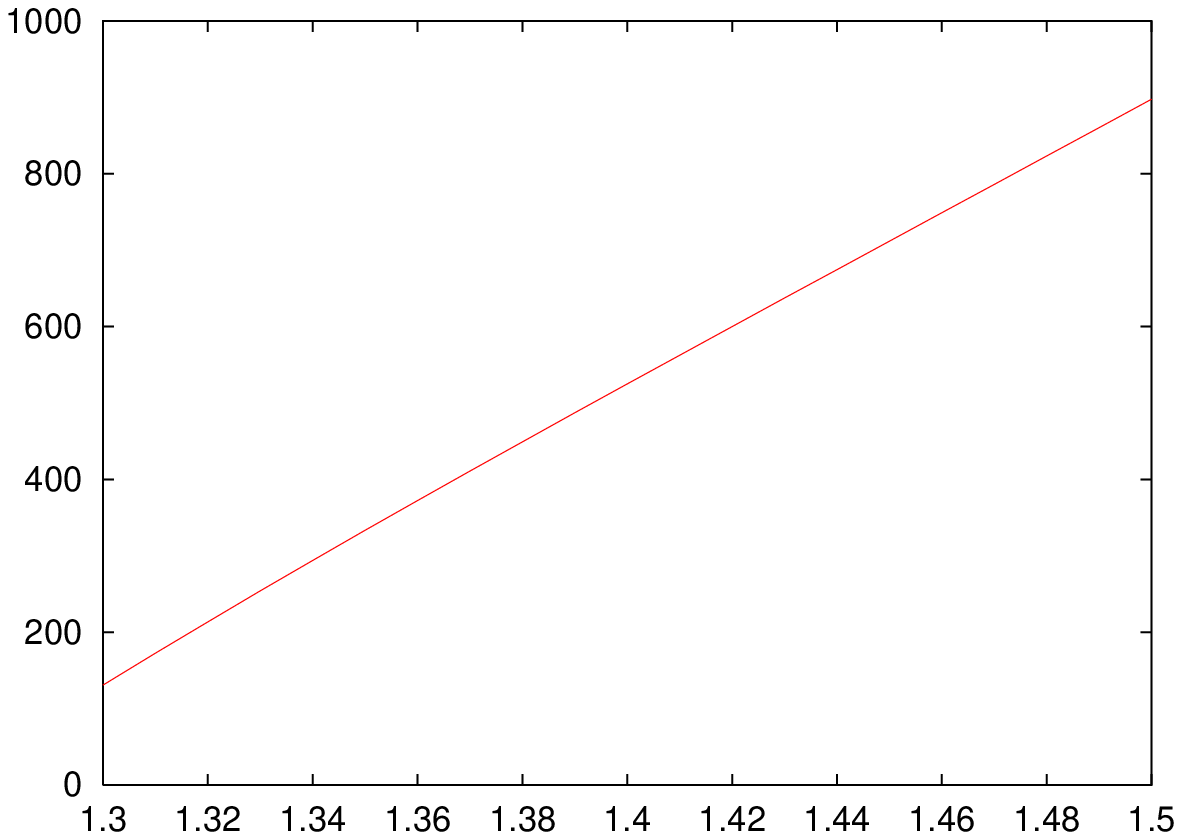}
\begin{picture}(0,0)
\setlength{\unitlength}{1mm}
\put(-11,60){$\Gamma_{\rho(2S) \to f_2\gamma}$}
\put(-8,55){(keV)}
\put(60,1){$m_{\rho({\rm 2S})}$~~(GeV)}
\end{picture}
\end{minipage}
\ec
\caption{Variation of the width for the decay $\rho(2S) \to f_2\gamma$ as a 
function of the mass of the $\rho(2S)$.}
\lb{f2gamma}
\efig

\subsection{Scalar mesons and glueballs}

In tables 4 and 5 we assumed that there is no mixing among the scalars, so
that the $f_0(1370)$ is pure $n\bar n$ and the $f_0(1710)$ is pure $s\bar s$.
The result of the mixing is that the bare $n\bar n$ and $s\bar s$ states 
contribute in varying degrees to each of the $f_0(1370)$, $f_0(1500)$, and
$f_0(1710)$. The variation of the radiative decay width of the $\rho(1700)$
and $\phi(1900)$ as a function of the mass of the $f_0$
are shown in figures \ref{f0nnbar} and \ref{f0ssbar} respectively.

\bfig
\bc
\begin{minipage}{90mm}
\epsfxsize90mm
\epsffile{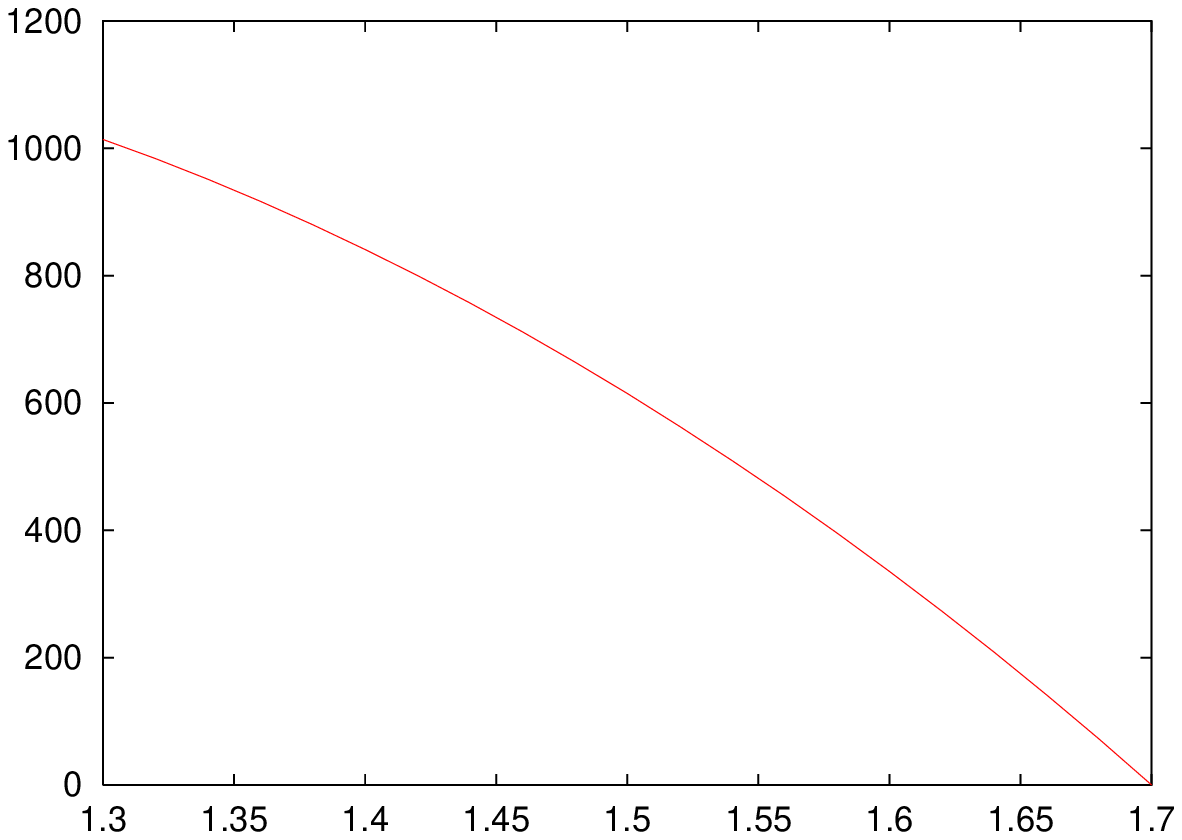}
\begin{picture}(0,0)
\setlength{\unitlength}{1mm}
\put(-12,60){$\Gamma_{\rho(1700) \to f_0\gamma}$}
\put(-8,55){(keV)}
\put(60,1){$m_{f_0}$~~(GeV)}
\end{picture}
\end{minipage}
\ec
\caption{Variation of the width for the decay $\rho(1700) \to f_0\gamma$ 
as a function of the mass of the $f_0$.}
\lb{f0nnbar}
\efig
\bfig
\bc
\begin{minipage}{90mm}
\epsfxsize90mm
\epsffile{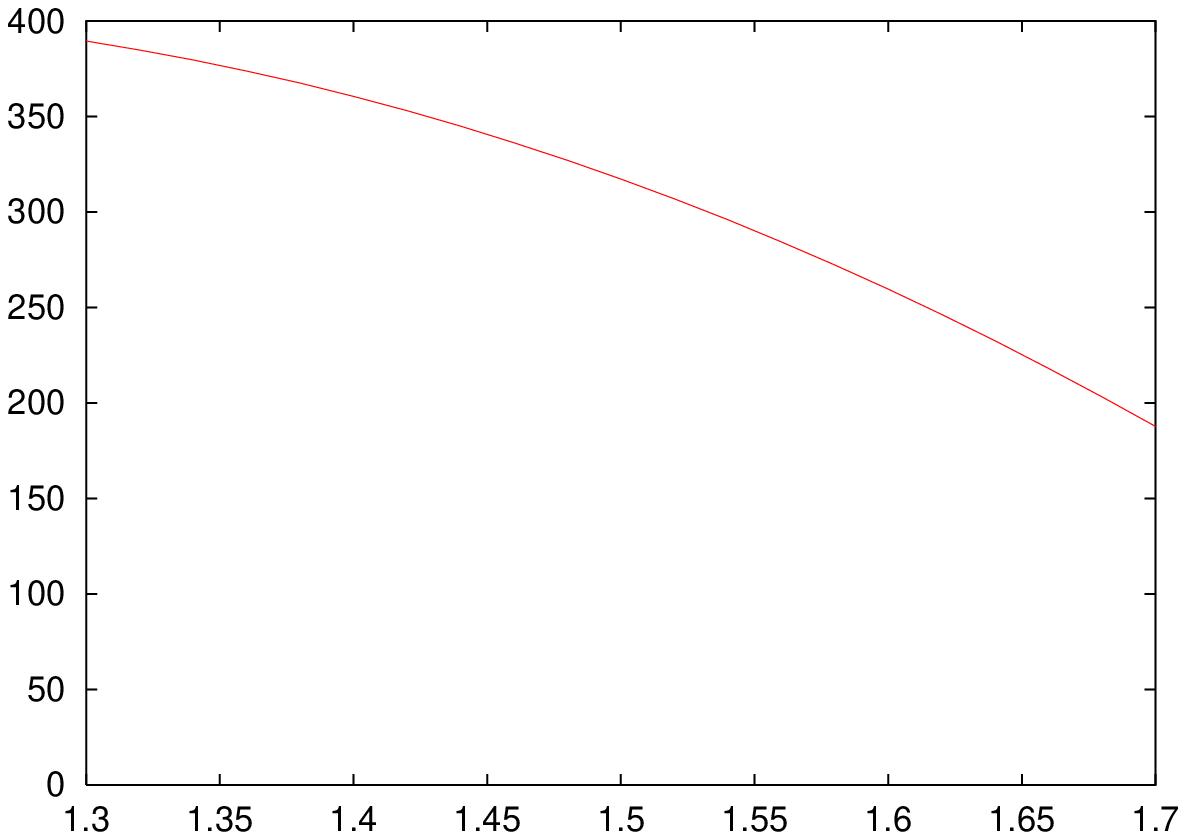}
\begin{picture}(0,0)
\setlength{\unitlength}{1mm}
\put(-13,61){$\Gamma_{\phi(1900) \to f_0\gamma}$}
\put(-8,56){(keV)}
\put(60,1){$m_{f_0}$~~(GeV)}
\end{picture}
\end{minipage}
\ec
\caption{Variation of the width for the decay $\phi(1900) \to f_0\gamma$ 
as a function of the mass of the $f_0$.}
\lb{f0ssbar}
\efig

Three different mixing scenarios have been proposed: the bare glueball is 
lighter than the bare $n\bar n$ state (the light glueball solution); the 
mass of the bare glueball is between the bare $n\bar n$ state and 
the bare $s\bar s$ state (the middleweight glueball solution); and the mass 
of the bare glueball is greater than the mass of the bare $s\bar s$ state.
The first two solutions have been obtained in \cite{CK01} and the third
has been suggested in \cite{LW00}. The effects of the mixing on the 
radiative decay widths of the $\rho(1700)$ and the $\phi(1900)$ to the 
three $f_0$ states are given in table 7 for each of these three cases.
The relative rates of the radiative decays of the $\rho(1700)$ to $f_0(1370)$ 
and $f_0(1500)$ change radically according to the presence
of the glueball admixture. So for a light glueball the decay to $f_0(1370)$
is relatively suppressed whereas for a heavy glueball it is substantial.
By contrast the effect on the decay to $f_0(1500)$ goes the other way.
Further, the $\phi(1900)$ would give a large width for the decay to
$f_0(1500)$ for a heavy glueball, but essentially zero for a light one.
The $f_0(1710)$ will be prominent in the decays of the $\phi(1900)$ for
all but the heaviest glueball.
It is clear that these decays do provide an effective
flavour-filtering mechanism. 

\bc
\btab{|c|ccc|ccc|}
\hl
&&$\rho(1700)$&&&$\phi(1900)$&\\
&L&M&H&L&M&H\\
\hl
$f_0(1370)$&174&440&603&7&8&31\\
\hl
$f_0(1500)$&520&301&98&5&35&261\\
\hl
$f_0(1710)$&&&&173&156&17\\
\hl
\etab
\ec

Table 7. Effect of mixing in the scalar sector of the $1^3P_0$ nonet.
The radiative widths, in keV, are given for three different mixing 
scenarios as described in the text: light glueball (L), middleweight
glueball (M) and heavy glueball (H).

\bigskip

Further, identifying the appropriate mixing 
scheme gives insight into the underlying physics of glueballs.
The existing phenomenology from hadronic decays seems to favour a light
glueball. Essentially, if the decays of the ``bare'' glueball are
flavour-independent, then the observed flavour dependencies for the 
hadronic decays of the physical mesons require \cite{CK01} the glueball 
mass to be at the 
low end of the range preferred by quenched-lattice studies.

The resolution of the isoscalar-scalar problem is intimately connected
with the isovector-scalar problem. The existence of any $a_0$ other than
the $a_0(980)$ remains controversial. The different mixing schemes for the 
isoscalar-scalar
mesons give rather different values for the mass of the bare $n\bar n$ state.
This mass will be reflected in the mass of its isovector partner, the $a_0$.
We see from table 4 that the width for the decay $\omega(1650) \to a_0\gamma$
is large, and from table 6 that it is a well-defined decay. So this decay
can provide independent information on the existence and properties of
the $a_0(1450)$.

\subsection{Hybrid meson radiative transitions}

Radiative transitions between hybrid states in the mass range under 
consideration are essentially ones in which spin-flip is required.
As the $q\bar q$ pair is in a spin-singlet state for a $1^{--}$ hybrid
and in a spin-triplet state for $0^{-+}$, $1^{-+}$ and $2^{-+}$ hybrids
then, analogously to (\ref{S11}), for the transition $(1^{--})_H \to
(J^{-+})_H\gamma$  one has
\be 
\Gamma = {{2J+1}\over{3}}{{4}\over{3}}\alpha p {{E_B}\over{m_A}}
{{p^2}\over{m_q^2}}IF^2_H
\lb{hwidth}
\ee
where $F_H$ is a form factor similar to (\ref{FS1}), unknown but of the
order of unity. If the mass ordering is indeed $0^{-+} < 1^{-+} < 1^{--}
< 2^{-+}$, the radiative decays of hybrid vectors to hybrid pseudoscalars 
and to exotic hybrids should be present but with widths less than 50 keV,
the exact value depending on the phase-space. However if the mass ordering
is the reverse, then the transition $(0^{-+})_H \to (1^{--})_H\gamma$ is
possible. With $J=1$ in (\ref{hwidth}), then for the decay $\pi_H(1800) \to
\omega_H(1500)$ this gives a rather healthy width of about 300 keV. Thus
the decay $\pi(1800) \to \omega(1420)$ or $\omega(1650)$ is a potential
discriminator among various assignments both for $\pi(1800)$ and the 
two $\omega$ states. For example, if the $\pi(1800)$ is a $3^1S_0$ 
$q\bar q$ state, these radiative decays will be very strongly suppressed,
with a width $\le 1$ keV 
because of the orthogonality of the wave functions, unless the $\omega(1420)$
or $\omega(1650)$ is $3^3S_1$, which is highly unlikely. Equally, if the
$\pi(1800)$ is a $2^1S_0$ $q\bar q$ state, then these radiative decays will 
have a width of more than 1000 keV. Thus if a radiative width of several
hundred keV is found, then the two states must be siblings, which should 
be most natural for hybrids.

So we note a clear hierarchy. The radiative decay width is $O$(1 MeV) for 
$q\bar q$ states in the same spatial state ($2S$ to $2S$ or $3S$ to $3S$);
it is $O$(300 keV) for hybrid to hybrid; and it is $O$(1 keV) for $3S$ to $2S$.
Heuristically, the $2S$ to $2S$ is big because of phase space, the hybrid
likewise but reduced by the gluon-quark spin coupling while the $3S$ to $2S$
is killed by the orthogonality of the wave functions.

Radiative transitions between hybrids and $q\bar q$ mesons are the most 
uncertain. In constituent gluon models such transitions are highly suppressed
as it is necessary to remove the gluon and to rearrange the colour degrees 
of freedom. In the flux tube model the mechanism is less explicit, and it
has been argued \cite{isgur} that the flux tube is excited as readily as
a quark, with no extra suppression for radiative transitions between hybrids
and $q\bar q$ states. However, even with no such suppression, the radiative decay 
of hybrid vectors to $^3P_J$ $q\bar q$ states is small as the transition is
necessarily magnetic spin-flip and is also suppressed by phase space. This contrasts 
with the radiative decay of $q\bar q$ vector mesons to $^3P_J$ $q\bar q$ mesons 
which proceeds in leading order by an electric transition. 
We shall report on this issue elsewhere.

\section{Conclusions}

In table 8 we show the final states for the dominant decays listed in tables
4 and 5. Obviously the experimentally cleanest signals come from the 
$\pi^+\pi^-$ decay of $f_2(1270)$ in $\rho^0(1450) \to f_2(1270)\gamma$ 
and the $K^+K^-$ decay of $f_2(1525)$ in $\phi(1690) \to f_2(1525)\gamma$.
The advantages are that the final-state mesons are comparatively narrow,
their decays are two-body and there are no neutrals in the final state
other than the photon. Additionally these are decays for which the calculation
is well-defined and the answers reliable, as we discussed in relation to 
table 6. An experimental check on the validity of the quark-model approach, 
apart from the total rate, is provided by the decay angular distribution.
From table 6 and (A.6) we see that we expect the electric-dipole term to
dominate and to give a nearly isotropic angular distribution 
$d\Gamma/d\cos\theta \propto 1+\cos^2\theta/13$. Any significant deviation 
from this would imply an unexpectedly-strong contribution from the 
magnetic-dipole term. These two decays are unique identifiers
of the $\rho(1450)$ and the $\phi(1690)$ respectively. As discussed in the 
previous section, radiative decays of a hybrid $\rho_H$ or $\phi_H$ to
these final states is strongly suppressed and so there is no ambiguity. 

\bc
\btab{|c|c|}
\hl
Decay&Final state\\
\hl
$\rho^0(1450) \to f_1(1285)\gamma$&$4\pi\gamma$, $\eta\pi\pi\gamma$\\
\hl
$\rho^0(1450) \to f_2(1270)\gamma$&$\pi\pi\gamma$\\
\hl
$\omega(1420) \to a_1(1260)\gamma$&$\pi^+\pi^-\pi^0\gamma$\\
\hl
$\omega(1420) \to a_2(1260)\gamma$&$\pi^+\pi^-\pi^0\gamma$\\
\hl
$\phi(1680) \to f_1(1420)\gamma$&$K{\bar K}\pi\gamma$\\
\hl
$\phi(1680) \to f_2(1525)\gamma$&$K{\bar K}\gamma$\\
\hl
$\phi(1900) \to f_1(1420)\gamma$&$K{\bar K}\pi\gamma$\\
\hl
\etab
\ec
\bc
Table 8. Final states for the dominant radiative decays of tables 4 and 5.
\ec

\bigskip

The decay $\phi(1900) \to f_1(1420)\gamma$ discriminates
between the $f_1(1420)$ and the $\eta(1440)$,
as can be seen from table 5. The nearness of the masses and widths of these
two states, and several common hadronic decay modes, have hitherto been 
sources of confusion. 
Although the magnitudes of the calculated radiative decays of the $1^3D_1$ 
states are subject to some uncertainty, the width of $\phi(1900) \to 
f_1(1420)\gamma$ is necessarily much larger than that of $\phi(1900) \to
\eta(1440)\gamma$. In the latter case the overlap of the wave functions
leads to a much stronger suppression of the decay than in the former case,
in any model. 

The relative rates of the radiative decays of the $\rho(1700)$ to $f_0(1370)$ 
and $f_0(1500)$, and of the $\phi(1900)$ to $f_0(1500)$ and $f_0(1710)$,
change radically according to the particular model for $q\bar q$-glueball 
mixing. 
The differences are sufficiently great, as can be
seen in table 7, for the appropriate mixing scheme to be identified and the
glueball mass determined. The physics that can be extracted from these decays
is significant and merits every effort to overcome the experimental problem
posed by the multiplicity of hadronic decay channels of the scalars.  

The observation of the decay $\omega(1650) \to a_0(1450)\gamma$ is of high 
priority as it establishes consistency between the masses of the isoscalar and
isovector $n\bar n$ states, with implications for the scalar glueball mass.
Fortunately the $a_0(1450)$ has comparatively simple decay modes, such as
$K\bar K$.

The larger partial widths should be measurable at the new high-intensity 
facilities being proposed. In some cases they may be measurable in the data 
from present experiments. We give two specific examples. The $\omega\eta$ 
decay of the $\omega(1650)$ has been observed in the E852 experiment 
\cite{E852}. If the $\omega(1650)$ is the $1D$ $q\bar q$ excitation of the 
$\omega$, then the $^3P_0$ model gives the partial width for this decay
as 13 MeV \cite{BCPS97}. The partial width for the radiative decay 
$\omega(1650) \to a_1(1260)\gamma$ is of the order of 1 MeV, that is about 
$8\%$ of the $\omega\eta$ width. The E852 experiment has several thousand 
events in the $\omega\eta$ channel, so we may expect several hundred events 
in the $a_1\gamma$ channel. Similarly both the $\rho(1450)$ and $\rho(1700)$ 
are seen by the VES collaboration \cite{VESc} in the $\rho\eta$ channel with 
several thousand events. Both these states have strong radiative decays, the
$\rho(1450)$ to $f_2(1270)\gamma$ and the $\rho(1700)$ to $f_1(1285)\gamma$. 
Assuming that the $\rho(1450)$ and $\rho(1700)$ are respectively the $2S$ and 
$1D$ excitations of the $\rho$, then the $^3P_0$ model gives the partial 
widths for the $\rho\eta$ decays of the $\rho(1450)$ and $\rho(1700)$ as 23 
MeV and 25 MeV respectively, so the radiative decays could again be present 
at the level of a few hundred events. 

\bigskip

\section{\bf Acknowledgements}

This work is supported, in part, by grants from the Particle Physics and
Astronomy Research Council, RFBR 00-15-96786, INTAS-RFBR 97-232 and the
European Community Human Mobility Program Eurodafne NCT98-0169.

\appendix
\section{Angular Distributions}
\setcounter{equation}{0}
\renewcommand{\theequation}{A.\arabic{equation}}

In the following equations the isospin factors $I$ are given in (\ref{iso1})
and (\ref{iso2}), $\lambda$ in (\ref{lambda}) and $\beta$ in (\ref{beta}).
The form factors $F_{S1}$, $F_{S2}$, $F_{D1}$ and $F_{D2}$ are defined
in (\ref{FS1}), (\ref{FS2}), (\ref{FD1}) and (\ref{FD2}) and the functions $G_E$
and $G_M$ in (\ref{GE}) and (\ref{GM}). 

${\bf 1^3S_1 \ra 1^1S_0}$
\be
\frac{d\Gamma^0_S}{d\cos\theta} = \frac{1}{2}\alpha
p\frac{E_B}{m_A}\frac{p^2}{m_q^2}(1+\cos^2\theta)IF^2_{S1}
\ee

${\bf 2^3S_1 \ra 1^1S_0}$
\be
\frac{d\Gamma}{d\cos\theta} = \frac{3}{2}
\left(\Big(\frac{\beta^2}{\beta_A^2}-1\Big)+\frac{2\lambda^2p^2}{3\beta_A^2}
\right)^2\frac{d\Gamma^0_S}{d\cos\theta}
\ee

${\bf 2^3S_1 \rightarrow2^1S_0}$
\be
\frac{d\Gamma}{d\cos\theta} = \frac{9}{4}
\left(\Big(\frac{5\beta^4}{3\beta_A^2\beta_B^2}-1\Big)+
\frac{4\lambda^2p^2}{3\beta^2}
\Big(\frac{\beta^4}{3\beta_A^2\beta_B^2}-1\Big)+\frac{4\lambda^2p^4}
{9\beta_A^2\beta_B^2} \right)^2\frac{d\Gamma^0_S}{d\cos\theta}
\ee

${\bf 2^3S_1 \rightarrow 1^3P_0}$
\beqa
\frac{d\Gamma}{d\cos\theta}&=&\frac{2}{9}\alpha p\frac{E_B}{m_A}
\frac{\beta^2}{m_q^2}(1+\cos^2\theta)\Big(1 - \frac{p^2}{2\beta^2}\Big)
\Big(G_E^2 - \frac{p^2}{2\beta^2}G_EG_M\nn\\
&&~~~~~~~ + \frac{1}{16}\Big(\frac{p^2}{\beta^2}\Big)^2G_M^2\Big)IF^2_{S2}
\eeqa

${\bf 2^3S_1 \rightarrow 1^3P_1}$
\beqa
\frac{d\Gamma}{d\cos\theta}&=&\alpha p\frac{E_B}{m_A}
\frac{\beta^2}{m_q^2}\Big((1-{\textstyle{\frac{1}{3}}}\cos^2\theta)
G_E^2-\frac{p^2}{3\beta^2}(1-\cos^2\theta)G_EG_M\nn\\
&&~~~~~~~~+\frac{1}{24}\Big(\frac{p^2}{\beta^2}\Big)^2(1-\cos^2\theta)G_M^2\Big)
IF^2_{2S}
\eeqa

${\bf 2^3S_1 \rightarrow 1^3P_2}$
\beqa
\frac{d\Gamma}{d\cos\theta}&=&\frac{13}{9}\alpha p\frac{E_B}{m_A}
\frac{\beta^2}{m_q^2}\Big((1+{\textstyle{\frac{1}{13}}}\cos^2\theta)G_E^2 +
\frac{2p^2}{13\beta^2}(2-\cos^2\theta)G_EG_M\nn\\
&&~~~~~~~~+\frac{1}{104}\Big(\frac{p^2}{\beta^2}\Big)^2(5-3\cos^2\theta)G_M^2
\Big)IF^2_{2S}
\eeqa

${\bf 1^3D_1 \ra 1^1S_0}$
\be
\frac{d\Gamma^0_D}{d\cos\theta} = \frac{1}{15}\alpha p\frac{E_B}{m_A}
\lambda^4\frac{p^4}{\beta^4}\frac{p^2}{m_q^2}(1+\cos^2\theta)IF^2_{D1}
\ee

${\bf 1^3D_1 \ra 2^1S_0}$
\be
\frac{d\Gamma}{d\cos\theta} = \frac{3}{2}\left(\Big(\frac{2\lambda^2p^2}{3\beta_B^2}-1
\Big)+\frac{7\beta^2}{3\beta_B^2} \right)^2\frac{d\Gamma^0_D}{d\cos\theta}
\ee

${\bf 1^3D_1 \ra 1^3P_0}$
\beqa
\frac{d\Gamma}{d\cos\theta} &=& \frac{10}{9}\alpha p\frac{E_B}{m_A}
\frac{\beta^2}{m_q^2}(1+\cos^2\theta)\Big(1+\frac{p^2}{5\beta^2}\lambda
(1+2\lambda)\nn\\
&&~~~~+\frac{p^4}{5\beta^4}\lambda^2(-1+2\lambda)\Big)^2IF^2_{D2}
\eeqa

${\bf 1^3D_1 \ra 1^3P_1}$
\beqa
\frac{d\Gamma}{d\cos\theta} &=& \frac{5}{6}\alpha p\frac{E_B}{m_A}
\frac{\beta^2}{m_q^2}\Big(\frac{1}{2}(3-\cos^2\theta)
+\frac{p^2}{5\beta^2}\lambda(11-5\cos^2\theta+
2\lambda(3-\cos^2\theta)\nn\\
&+&\frac{p^4}{50\beta^4}\lambda^2
(1+17\cos^2\theta+4\lambda(31-25\cos^2\theta)+4\lambda^2(3-
\cos^2\theta))\nn\\
&+&\frac{8p^6}{25\beta^6}\lambda^3(1-\cos^2\theta)(-2
+3\lambda+2\lambda^2)\nn\\
&+&\frac{4p^8}{25\beta^8}\lambda^4
(1-\cos^2\theta)(1-4\lambda+4\lambda^2)\Big)IF^2_{D2}
\eeqa

${\bf 1^3D_1 \ra 1^3P_2}$
\beqa
\frac{d\Gamma}{d\cos\theta} &=& \frac{1}{15}\alpha p\frac{E_B}{m_A}
\frac{\beta^2}{m_q^2}\Big(\frac{1}{12}(13+
\cos^2\theta)\nn\\
&+&\frac{p^2}{6\beta^2}\lambda(11+35\cos^2\theta)
+2\lambda(17-43\cos^2\theta)\nn\\
&+&\frac{p^4}{12\beta^4}\lambda(83(1+\cos^2\theta)-
4\lambda(103+7\cos^2\theta)+20\lambda^2(41-19\cos^2\theta))\nn\\
&+&\frac{p^6}{3\beta^6}\lambda^3(17-7\cos^2\theta)
-12\lambda(9-5\cos^2\theta)+4\lambda^2(37-23\cos^2\theta))\nn\\
&+&\frac{p^8}{3\beta^8}\lambda^4(7-5\cos^2\theta)(1
-4\lambda+4\lambda^2)\Big)IF^2_{D2}
\eeqa

\section{Radiative Helicity Amplitudes}
\setcounter{equation}{0}
\renewcommand{\theequation}{B.\arabic{equation}}

In this appendix we show how the amplitudes used in the main body
of this paper are related to helicity amplitudes that appear
elsewhere in the literature. This also enables application of some
of our results, which have been specified for $e^+e^-$ annihilation,
to be taken over to photoproduction. 

Following \cite{CKO69,qpartons}
the electromagnetic interaction may be written in the form
\be
J_{em} = \sum_{j=1}^{2} e_j \mu \Big( -2i\vec{s} \cdot \vec{p} \times \vec{A}
- g^{-1}(\vec{k} + \vec{k'}) \cdot \vec{A}\Big)
\lb{B1}
\ee
where $e_j$ is the quark charge in units of $e$, $\mu \equiv 
g\sqrt{\alpha/2m_q} = 0.13$GeV is
the quark-scaled magnetic moment
and $\vec{A} = \vec{\epsilon} \sqrt{4\pi/2\omega}~e^{i\vec{p}\cdot
\vec{r_j}}$. Choosing the $\hat{z}$ axis to be along the photon momentum,
and $\vec{\epsilon} = - \frac{1}{\sqrt{2}}(1,i,0)$ for $J_z=+1$ then 
(\ref{B1}) becomes
\be
J_{em} = 2\sqrt{\frac{\pi}{\omega}} \mu
\sum_{j=1}^{2} e_j \Big( p (S_x + i S_y) + g^{-1}(k_x + ik_y)\Big)e^{ipz}
\ee
to be compared with (7.3) in \cite{qpartons}, whereby
after summing over two quarks the matrix element becomes
\be
4\sqrt{\frac{\pi}{\omega}} \mu \langle e \rangle
 ( p \langle\sigma_{+}\rangle R_{L0} +g^{-1}
\langle L_+ \rangle R_{L1})
\ee
where $\langle e \rangle^2 \equiv I$ of (\ref{rate}) and
\beqa
R_{L1} &\equiv& \langle \psi_{L1}^* | e^{ipr/2}(-i){{d}\over{dr}}|\psi_{00} \rangle 
\nonumber \\
R_{L0} &\equiv& \langle \psi_{L0}^* | e^{ipr/2} | \psi_{00} \rangle
\eeqa
where the $\psi_{L,L_z}$ are the bound-state wave functions for the for the 
$q\bar q$ state with orbital angular momentum $L,L_z$ \cite{CKO69}.

Radiative widths are then
\be
\frac{\Gamma(A\to B \gamma)}{8m_{A}E_{B}}
= \frac{p}{8\pi m_A^2}\frac{2J_B+1}{2J_A + 1}\langle e\rangle^2
\sum_{\lambda}|M_{\lambda}|^2
\ee
and $\omega \equiv |p|$ for real photons;
$\langle e \rangle = \textstyle{{1}\over{2}}$ for
$I=0 \to I=1$ and $\langle e \rangle =\textstyle{{1}\over{6}}$ for 
$I=0(1) \to I=0(1)$ and the $M_\lambda$ are helicity amplitudes with
helicity $\lambda$.

\subsection{M1 transitions}

As an illustration and check on the normalisation we compute 
$\Gamma(\omega \to \pi \gamma)$.

The matrix element becomes ($\langle\sigma_+\rangle = 1/\sqrt{2}$)
\be
M_{+1} = \mu \sqrt{2\pi p} R_{L0}
\ee
for $J_z^{\gamma}=+1$ and $\langle e \rangle = 1/2$. Hence
\beqa
\Gamma(\omega \to \pi \gamma)
&=& \frac{4}{3} \mu^2 p^3 \frac{E_{\pi}}{m_{\omega}}
| R_{00}(p) |^2 \nonumber \\
&\to& \frac{4}{3} \alpha p \frac{E_{\pi}}{m_{\omega}} \frac{g p^2}{m_q^2} 
\frac{1}{4}| R_{00}(p) |^2
\eeqa
to be compared with (\ref{S11}) with $m_q = 0.33$, $g = 1$. 
In the harmonic oscillator basis 
\be
R_{00}(p) 
\equiv \exp\Big(-\frac{p^2}{16 \beta^2}\Big) 
\ee
which is the $\beta_A = \beta_B$ limit of 
$F_{S1}$ of (\ref{FS1}). Hence if $\beta \sim 0.4$GeV 
\cite{KI87,ABS96,BCPS97}, then
\be
\Gamma(\omega \to \pi \gamma)  
\sim 0.6 MeV
\ee
In the text we show that this value for $\beta$ does not fit well with 
the detailed spectroscopy, and when realistic values are used, the actual
width for $\omega \to \pi \gamma$ is somewhat reduced. See figure 1.

An analogous calculation for the radiative transition from
the $\omega(2^3S_1)$ involves a radial wavefunction and
different magnitude for $p$, $p^*$ say. Thus

\be
\frac{\Gamma(\omega(2^3S_1) \to \pi \gamma)}{\Gamma(\omega \to \pi \gamma)}
= \Big(\frac{p^*}{p}\Big)^3 \frac{1}{6} \Big(\frac{p^{*2}}{8\beta^2}\Big)^2
\exp\Big(\frac{p^2 - p^{*2}}{8 \beta^2}\Big)
\ee

This also sets the scale for transitions between
$\pi(1800)$ and $\omega(1420/1650)$ which potentially bear on the question
of hybrid states.

\subsection{Application to excited states: E1 and M2 transitions}

Helicity amplitudes follow from 
the most general form of the single quark
interaction with a transversely polarised photon in the algebraic
form \cite{qpartons}
\be
J^{em}_+ \equiv (AL_+ + B\sigma_+ + C\sigma_zL_+ + D\sigma_-L_{++})4\mu
\sqrt{\pi p}\langle e \rangle
\ee
In the non-relativistic limit $C \equiv D =0$ and
\beqa
A\langle L_+\rangle &\equiv&  R_{L1}/pg \nonumber \\
B\langle\sigma_+\rangle &\equiv&   R_{L0}/\sqrt{2}
\eeqa

We will be particularly interested in transitions between $2S$ and $1P$ 
states. In the Gaussian wave function approach we have
\beqa 
A\langle L_+\rangle  &\equiv&   R_{L1}/pg\nn\\
&=& \frac{i\beta \sqrt{2}}{pg\sqrt{3}} \Big(1+ \frac{p^2}{16\beta^2}\Big)
\exp\Big(\frac{-p^2}
{16\beta^2}\Big)
\equiv \frac{i\beta\sqrt{2}}{gp\sqrt{3}} G_E \nonumber \\
B\langle\sigma_+\rangle &\equiv& R_{L0}/\sqrt{2}\nn\\
&=& \frac{ip}{\beta2\sqrt{6}}\Big(1-\frac{p^2}{16\beta^2}\Big)  
\exp\Big(\frac{-p^2}{16\beta^2}\Big)
\equiv \frac{i\beta\sqrt{2}}{p\sqrt{3}}(\frac{p^2}{4\beta^2})G_M
\eeqa

If we are only interested in terms up to $O(p^4)$ we can approximate the above
as
\beqa
A\langle L_+\rangle &\to& {{i\beta\sqrt{2}}\over{gp\sqrt{3}}}\nn\\
B\langle\sigma_+\rangle &\to& \frac{ip}
{\beta2\sqrt{6}}\Big(1-\frac{p^2}{8\beta^2}\Big)
\eeqa
The widths are then
\be
\Gamma(\rho(2S) \to f_J \gamma) = (2J+1) \frac{8}{3} \Big(\frac{\mu}{g}
\Big)^2 p^3 
\frac{E_j}{m_{\rho}} 
\sum_{\lambda\geq 0}|M_{\lambda}|^2
\ee
where $\lambda$ refers to the helicity of the state and the matrix elements 
$|M_{\lambda}|$ are:
\beqa
f_0: M_0 &=& (A-B)/\sqrt{3} \nn\\
f_1: M_0 &=& A/\sqrt{2}\nn\\
     M_1 &=& (A-B)/\sqrt{2} \nn\\
f_2: M_0 &=& (A+2B)/\sqrt{6}\nn\\
     M_1 &=& (A+B)/\sqrt{2}\nn\\
     M_2 &=& A
\eeqa
As a specific application,
\beqa
\Gamma(\rho(2S) \to f_0 \gamma) &=&  \frac{8}{3} \Big(\frac{\mu}{g}\Big)^2 p^3 
\frac{E_{f_0}}{m_{\rho}}
|(A-B)/\sqrt{3}|^2 \nonumber \\
&=& \frac{4}{27} \frac{\alpha}{m_q^2} p\beta^2 \frac{E_{f_0}}{m_{\rho}}  
\bigg(\Big(1+\frac{p^2}{16\beta^2}\Big) - g \frac{p^2}{4\beta^2 }\bigg)^2 
\exp\Big(-{{p^2}\over{8\beta^2}}\Big)\nn\\
&=&  \frac{4}{27} \frac{\alpha}{m_q^2} p\beta^2 
\frac{E_{f_0}}{m_{\rho}} 
\Big(G_E - \frac{p^2}{4\beta^2}G_M\Big)^2  F^2
\eeqa
using $\mu/g \equiv \sqrt{\alpha}/2m_q$.
The form in (\ref{S21}) reduces to this in the
particular case where all $\beta$ values are the same and 
the isospin factor $I= 1/4$

Note also the following translations:
\beqa
G_E &\equiv& \sqrt{\frac{3}{2}} \frac{p}{\beta} A\nn\\
G_M &\equiv& \sqrt{\frac{3}{2}} \frac{p}{\beta} \frac{p^2}{4\beta^2} B
\eeqa

\end{document}